\documentclass{emulateapj}
\usepackage{apjfonts}

\lefthead{JUNG ET AL.}
\righthead{ }

\begin{document}
\title{Binary Source Microlensing Event OGLE-2016-BLG-0733: 
Interpretation of A Long-term Asymmetric Perturbation}

\author{
Y.~K.~Jung$^{1,22}$, 
A.~Udalski$^{2,23}$, 
J.~C.~Yee$^{1,22}$,
T.~Sumi$^{3,24}$ \\
and \\
A.~Gould$^{4,5,6}$, C.~Han$^{7}$, M.~D.~Albrow$^{8}$, C.-U.~Lee$^{4,9}$, S.-L.~Kim$^{4,9}$, 
S.-J.~Chung$^{4,9}$, K.-H.~Hwang$^{4}$, Y.-H.~Ryu$^{4}$, I.-G.~Shin$^{1}$, W.~Zhu$^{5}$, 
S.-M.~Cha$^{4,10}$, D.-J.~Kim$^{4}$, Y.~Lee$^{4,10}$, B.-G.~Park$^{4,9}$, R.~W.~Pogge$^{5}$ \\ 
(The KMTNet Collaboration) \\
P.~Pietrukowicz$^{2}$, S.~Koz{\l}owski$^{2}$, R.~Poleski$^{2,5}$, J.~Skowron$^{2}$,
P.~Mr{\'o}z$^{2}$, M.~K.~Szyma{\'n}ski$^{2}$, I.~Soszy{\'n}ski$^{2}$, M.~Pawlak$^{2}$,   
K.~Ulaczyk$^{2}$ \\
(The OGLE Collaboration) \\
F.~Abe$^{11}$, D.~P.~Bennett$^{12,13}$, R.~Barry$^{14}$, I.~A.~Bond$^{15}$, Y.~Asakura$^{11}$, 
A.~Bhattacharya$^{12}$,M.~Donachie$^{16}$, M.~Freeman$^{16}$, A.~Fukui$^{17}$, 
Y.~Hirao$^{3}$, Y.~Itow$^{11}$, N.~Koshimoto$^{3}$, M.C.A.~Li$^{16}$, 
C.H.~Ling$^{18}$, K.~Masuda$^{11}$, Y.~Matsubara$^{11}$, Y.~Muraki$^{11}$, 
M.~Nagakane$^{3}$, H.~Oyokawa$^{11}$, N.~J.~Rattenbury$^{16}$, A.~Sharan$^{16}$, 
D.J.~Sullivan$^{19}$, D.~Suzuki$^{13}$, P.~J.~Tristram$^{20}$, T.~Yamada$^{3}$, 
T.~Yamada$^{21}$, A.~Yonehara$^{21}$ \\ 
(The MOA Collaboration)
}

\bigskip\bigskip
\affil{$^{1}$Smithsonian Astrophysical Observatory, 60 Garden St., Cambridge, MA 02138, USA}
\affil{$^{2}$Warsaw University Observatory, Al. Ujazdowskie 4, 00-478 Warszawa, Poland}
\affil{$^{3}$Department of Earth and Space Science, Graduate School of Science, Osaka University, Toyonaka, Osaka 560-0043, Japan}
\affil{$^{4}$Korea Astronomy and Space Science Institute, Daejon 305-348, Republic of Korea}
\affil{$^{5}$Department of Astronomy, Ohio State University, 140 W. 18th Ave., Columbus, OH 43210, USA}
\affil{$^{6}$Max-Planck-Institute for Astronomy, K$\rm \ddot{o}$nigstuhl 17, 69117 Heidelberg, Germany}
\affil{$^{7}$Department of Physics, Chungbuk National University, Cheongju 371-763, Republic of Korea}
\affil{$^{8}$University of Canterbury, Department of Physics and Astronomy, Private Bag 4800, Christchurch 8020, New Zealand}
\affil{$^{9}$Korea University of Science and Technology, 217 Gajeong-ro, Yuseong-gu, Daejeon 34113, Korea}
\affil{$^{10}$School of Space Research, Kyung Hee University, Yongin 446-701, Republic of Korea}
\affil{$^{11}$Institute for Space-Earth Environmental Research, Nagoya University, Nagoya 464-8601, Japan}
\affil{$^{12}$Department of Physics, University of Notre Dame, Notre Dame, IN 46556, USA}
\affil{$^{13}$Laboratory for Exoplanets and Stellar Astrophysics, NASA/Goddard Space Flight Center, Greenbelt, MD 20771, USA}
\affil{$^{14}$Astrophysics Science Division, NASA Goddard Space Flight Center, Greenbelt, MD 20771, USA}
\affil{$^{15}$Institute of Natural and Mathematical Sciences, Massey University, Auckland 0745, New Zealand}
\affil{$^{16}$Department of Physics, University of Auckland, Private Bag 92019, Auckland, New Zealand}
\affil{$^{17}$Okayama Astrophysical Observatory, National Astronomical Observatory of Japan, 3037-5 Honjo, Kamogata, Asakuchi, Okayama 719-0232, Japan}
\affil{$^{18}$Institute of Information and Mathematical Sciences, Massey University, Private Bag 102-904, North Shore Mail Centre, Auckland, New Zealand}
\affil{$^{19}$School of Chemical and Physical Sciences, Victoria University, Wellington, New Zealand}
\affil{$^{20}$Mt. John University Observatory, P.O. Box 56, Lake Tekapo 8770, New Zealand}
\affil{$^{21}$Department of Physics, Faculty of Science, Kyoto Sangyo University, 603-8555 Kyoto, Japan}
\footnotetext[22]{The KMTNet Collaboration.}
\footnotetext[23]{The OGLE Collaboration.}
\footnotetext[24]{The MOA Collaboration.}

\begin{abstract}
In the process of analyzing an observed light curve, one often 
confronts various scenarios that can mimic the planetary signals 
causing difficulties in the accurate interpretation of the lens system. 
In this paper, we present the analysis of the microlensing event 
OGLE-2016-BLG-0733. The light curve of the event shows 
a long-term asymmetric perturbation that would appear to be due 
to a planet. From the detailed modeling of the lensing light curve, 
however, we find that the perturbation originates from the binarity 
of the source rather than the lens. This result demonstrates 
that binary sources with roughly equal-luminosity components can mimic 
long-term perturbations induced by planets with projected separations 
near the Einstein ring. The result also represents the importance of 
the consideration of various interpretations in planet-like perturbations 
and of high-cadence observations for ensuring the unambiguous 
detection of the planet.
\end{abstract}


\section{Introduction}

Since its proposal as a means to prove the mass distribution of compact dark 
objects in the halo of the Galaxy \citep{paczynski86,alcock93,udalski93}, 
microlensing has been applied to various fields of astronomy 
including planetary science \citep{mao91,gould92b,bond04}. 
Although the number of planets detected by the microlensing method is far less 
than that of planets discovered by other major methods such as the transit 
and radial velocity methods, microlensing provides a unique tool to 
find planets that are hard to detect by other techniques, such as planets 
around faint and dark objects, planets located at or beyond the snow line, 
and planets not gravitationally bound to their host stars (see a detailed 
review by \citealt{gaudi12}). Therefore, it is complementary to other methods, 
enabling the comprehensive study of extrasolar planets. 
Furthermore, with the advent of second generation microlensing surveys 
such as the Microlensing Observations in Astrophysics \citep[MOA-II:][]{sako08}, 
the Optical Gravitational Lensing Experiment \citep[OGLE-IV:][]{udalski15}, and
the Korea Microlensing Telescope Network \citep[KMTNet:][]{kim16}, and
future microlensing surveys in space \citep[e.q., {\it WFIRST}:][]{spergel15}, 
we are on the brink of large numbers of planet discoveries, 
comparable to those found by the transit and radial velocity methods.

The microlensing signal of a planet is almost always a short-term perturbation 
on the single-lens light curve caused by the host of the planet. The signal is 
usually produced by the passage of the source star close to or over a caustic, 
which are the locations on the source plane where the lens-mapping equation 
would be singular, and thus the lensing magnification of a point source would 
be infinite \citep{schneider86}. The shape, size, and number of closed caustic 
curves are characterized by the mass ratio and the separation between the host 
and the planet \citep{erdl93}. As a result, planetary perturbations take various 
forms depending on the characteristics of the planetary system as well as the 
source trajectory.

Interpretation of the planetary microlensing signal and the characterization of 
the planetary system require detailed analyses of the signal in the observed light 
curve. However, accurate interpretation of the signal is often hampered by 
various scenarios that can mimic the planetary signals. For example, 
the signals induced from two different binary-lens systems can be similar 
despite the great difference between their underlying physical characteristics. 
This degeneracy was presented by \citet{choi12} and \citet{bozza16}, 
where they showed that short-term anomalies of a subset of lensing light curves 
could be explained equally well by either a binary lens with roughly equal mass 
components or a binary with a very large mass ratio (i.e., planetary system). 
In addition, \citet{gould13} showed that microlensing of star spots can 
give rise to light-curve deviations that very strongly mimic planetary anomalies. 
Furthermore, the binarity of the source star rather than the lens 
also can masquerade as planetary signals. \citet{gaudi98} pointed out that if 
a source is a binary star with a large flux ratio between the components and a 
single lens passes close to the binary source, the light curve 
can take a form of a standard single-lensing curve superposed by a short-term 
perturbation, which is similar to that of a planetary event. \citet{hwang13} 
showed that this degeneracy could be severe by presenting the light curve of an 
actually observed event.

Current and future microlensing surveys demand a uniform, en masse search for 
and analysis of planetary signals, a search that is expected to discover 
thousands of planets amongst tens of thousands of microlensing events 
\citep{henderson14,spergel15}. Ultimately, such systematic searches will need 
to account for the various potential sources of false positives in order to 
properly understand the planetary sample. In this paper, we present the 
analysis of OGLE-2016-BLG-0733, which could be mistaken for a planetary signal 
in such a large scale analysis. In Section 2, we describe the observation and 
the data reduction of the lensing event. In Section 3, we show that the long-term 
asymmetric perturbation in the light curve is caused by a binary source 
passing near the line of sight to a single lens and not due to a planet. 
In Section 4, we discuss the implications of the results.

\begin{figure}[th]
\epsscale{1.2}
\plotone{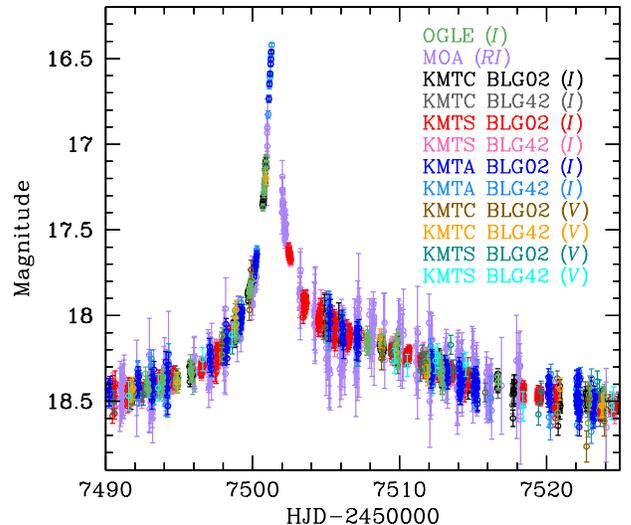}
\caption{\label{fig:one}
Light curve of OGLE-2016-BLG-0733. The notation after the each 
observatory represents the passband of observation.
}
\end{figure}

\section{Observation}

The equatorial and Galactic coordinates of the microlensing event 
OGLE-2016-BLG-0733 are $(\alpha,\delta)_{\rm J2000} = 
(17^{\rm h}55^{\rm m}54^{\rm s}\hskip-2pt.29, -29^{\circ}50'28''\hskip-2pt.2)$ 
and $(l,b) = (0.369^\circ, -2.388^\circ)$, respectively. 
The event was discovered on 2016 Apr 19 UT 19:16 by the Early Warning System 
\citep[EWS:][]{udalski15} of the OGLE survey, which is conducted using the 1.3m 
Warsaw telescope located at Las Campanas Observatory in Chile. The MOA survey 
independently discovered the event using the 1.8m telescope located at Mt. John 
Observatory in New Zealand. In the MOA lensing event list, 
it is labeled MOA-2016-BLG-202. 

The event was also in the footprint of the KMTNet survey. The KMTNet survey 
utilizes three identical 1.6m telescopes that are located at Cerro Tololo 
Interamerican Observatory in Chile (KMTC), South African Astronomical 
Observatory in South Africa (KMTS), and Siding Spring Observatory in 
Australia (KMTA). Each of KMTNet telescope is equipped with 
a $4.0~\rm deg^{2}$ camera. With these instruments, the KMTNet survey 
covers $(12, 40, 80)~\rm deg^{2}$ areas of the Galactic bulge with observation 
cadences of $(4, \geq1, \geq0.4)~\rm hr^{-1}$, respectively. For the three major 
fields covered by a $4~\rm hr^{-1}$ cadence, alternating observations cover 
the sky with a $6'$ offset in order to compensate the gap between camera chips. 
The event lay in one such pair of fields (BLG02 and BLG42) and thus the cadence 
of the event reached up to $4~\rm hr^{-1}$. Combining this feature with its 
globally distributed telescopes, the KMTNet survey densely and continuously 
covered the event.

Photometry data used for the analysis were processed using the 
customized pipelines that are developed by the individual survey 
groups: \citet{udalski03}, \citet{bond01}, and \citet{albrow09} 
for the OGLE, MOA, and KMTNet groups, respectively. These pipelines 
are based on the Difference Imaging Analysis method 
\citep[DIA:][]{alard98,wozniak00,albrow09}. 

\begin{figure}[th]
\epsscale{1.2}
\plotone{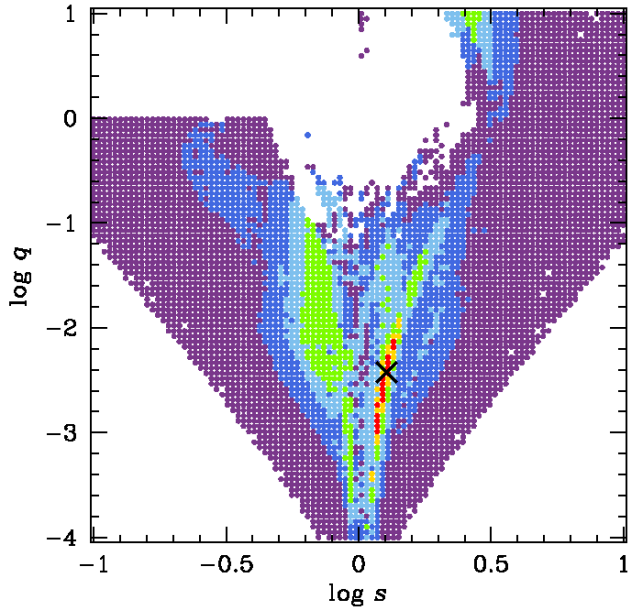}
\caption{\label{fig:two}
The $\Delta\chi^{2}$ surface of $(s, q)$ parameter space derived 
from the initial grid search. Each color represents the contour level of 
$\Delta\chi^{2} < 20\sigma$ (red), $< 40\sigma$ (yellow), 
$< 60\sigma$ (green), $< 80\sigma$ (light blue), $< 100\sigma$ (blue), 
and $<120{\sigma}$ (purple), respectively. The location of the 
best-fit solution is marked by the cross symbol. 
}
\end{figure}

For the case of KMTNet, data were additionally reduced using the pyDIA 
\footnote{http://www2.phys.canterbury.ac.nz/{\textasciitilde}mda45/pyDIA} 
software solely to determine the source color. pyDIA is a new modular python 
package for performing star detection, difference imaging and photometry 
on crowded astronomical images. For difference imaging, it uses the 
algorithm of \citet{bramich13} with extended delta basis functions 
that allows independent degrees of control for the differential 
photometric scaling as well as shape changes of the PSF between images. 
When available, it uses GPU hardware acceleration for the 
computationally intensive tasks of difference-imaging and PSF fitting.

A wealth of experience shows that while the error bars reported 
by microlensing photometry codes are, in general, monotonically related 
to the true errors, the transformation from one to the other varies 
from event to event and from observatory to observatory. To compensate 
for this effect, we renormalize the error bars of the individual data 
sets by following the method of \citet{yee12} with the equation   
\begin{equation}
\sigma^\prime_{i,j} = \sqrt{\sigma_{\rm min,\it j}^2 + (k_j\sigma_{i,j})^2},
\label{eq1}
\end{equation}
where $\sigma_{i,j}$ is the error on the $i$-th point from $j$-th observatory 
and $\sigma_{\rm min,\it j}$ and $k_j$ are the correction parameters for 
$j$-th observatory. For each data set, we first adjust $\sigma_{\rm min,\it j}$ 
in order to make the cumulative distribution function of $\chi^{2}$ sorted by 
lensing magnification linear. Then, we rescale the error bars using $k_j$ 
in order to make $\chi^{2}$ per degree of freedom ($\chi^2/{\rm dof}$) unity. 
Note that the normalization process is conducted based on the best-fit 
(binary-source) model and we remove $3\sigma$ outliers from the model 
during the process.

\section{Analysis}

Analysis of the lensing light curve is carried out by conducting modeling 
of the observed light curve to find the best-fit lensing parameters. 
As presented in Figure~\ref{fig:one}, the light curve 
of OGLE-2016-BLG-0733 shows deviations from the smooth and 
symmetric form of an event produced by a single lens. 
The deviations appear in two regions, one strong anomaly near 
the peak at ${\rm HJD}' (={\rm HJD}-2450000) \sim 7501.4$ 
and the other weak but long-term anomaly in the falling side 
of the light curve. Because such deviations are broadly consistent 
with a binary lens, and because identifying binary and 
especially planetary lens is a principal goal of micolensing 
studies, these anomalies triggered a search for binary-lens 
solutions even before the event returned to baseline. 
The predictions made by these ongoing ``real-time'' 
models were, broadly speaking, confirmed. Although the model 
nevertheless proves incorrect, we first study how it partially 
succeeds in explaining the data.

\begin{deluxetable}{lrr}
\tablecaption{Lensing Parameters\label{table:one}}
\tablewidth{0pt}
\tablehead{
\multicolumn{1}{l}{Parameters} &
\multicolumn{1}{c}{Binary-lens} &
\multicolumn{1}{c}{Binary-source}
}
\startdata
$\chi^2$/dof              &     12955.8/12347      &     12411.8/12347      \\
$t_{0,1}$ (${\rm HJD'}$)  & 7501.570 $\pm$ 0.004   &  7501.374 $\pm$ 0.005  \\
$t_{0,2}$ (${\rm HJD'}$)  & --                     &  7507.804 $\pm$ 0.116  \\
$u_{0,1}$                 &    0.013 $\pm$ 0.001   &     0.015 $\pm$ 0.001  \\
$u_{0,2}$                 & --                     &     0.365 $\pm$ 0.033  \\
$t_{\rm E}$ (days)        &   27.838 $\pm$ 0.847   &    14.428 $\pm$ 0.602  \\
$s$                       &    1.281 $\pm$ 0.007   & --                     \\
$q$ ($10^{-3}$)           &    3.912 $\pm$ 0.258   & --                     \\
$\alpha$ (rad)            &    0.051 $\pm$ 0.001   & --                     \\
$\rho_\ast$ ($10^{-3}$)   &    8.317 $\pm$ 0.392   & --                     \\
$q_{F,RI}$                 & --                     &    2.078 $\pm$ 0.201   \\
$q_{F,I}$                 & --                     &    1.853 $\pm$ 0.177   
\enddata 
\vspace{0.05cm}
\tablecomments{
${\rm HJD}'= {\rm HJD}-2450000$
}
\end{deluxetable}

\subsection{Binary-lens Interpretation}
Adopting the parametrization of \citet{jung15} for the description of 
a standard binary-lens light curve, we search for a binary-lens solution 
by following the method of \citet{jung15}, one of the well-established 
modeling procedure developed based on the map-making method \citep{dong06}. 
\footnote{In the modeling procedure, we adopt the limb-darkening coefficients 
$(\Gamma_{V}, \Gamma_{R}, \Gamma_{I}) = (0.74, 0.66, 0.57)$ from \citet{claret00} 
based on the source color measurement (see Section 3.3). 
For the MOA data, which are obtained from a non-standard
filter system, we use $\Gamma_{RI} = (\Gamma_{R} + \Gamma_{I})/2 = 0.62$.}
Figure~\ref{fig:two} shows the $\Delta\chi^{2}$ surface of $(s, q)$ parameter 
space derived from the initial grid search. Here $s$ is the projected binary 
separation (normalized to the angular Einstein radius of the lens system, 
$\theta_{\rm E}$) and $q$ $(= M_{2}/M_{1})$ is the mass ratio between the 
lens components, and they are divided into $100\times100$ grids in the range of 
$-1.0 < {\rm log}~s < 1.0$ and $-4.0 < {\rm log}~q < 1.0$, respectively. 
From the grid search, we find one local minimum. By refining the local 
solutions, it is found that the best-fit binary-lens model 
has two masses with a separation $s \sim 1.28$ and a mass ratio 
$q \sim 0.004$, which would make the companion a planetary-mass object 
projected near the Einstein radius of its host.

\begin{figure}[th]
\epsscale{1.2}
\plotone{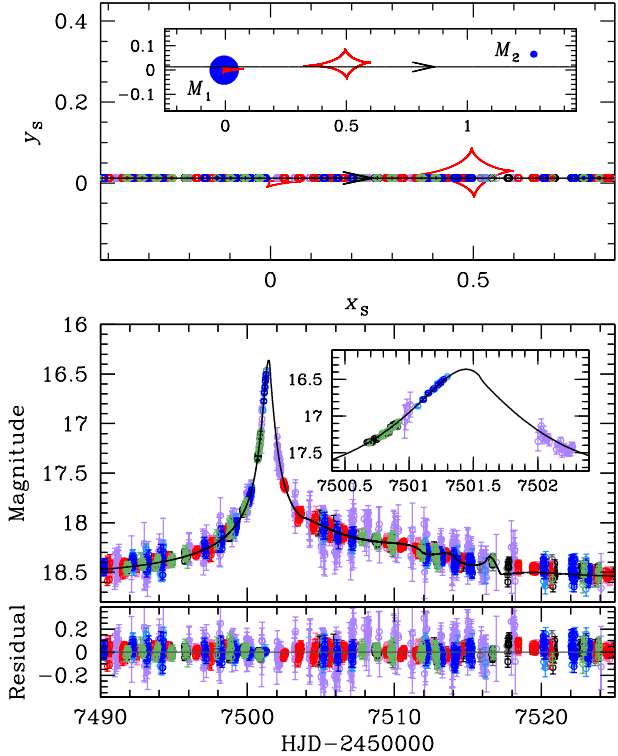}
\caption{\label{fig:three}
Geometry and light curve of the binary-lens model. 
(1) The upper panel shows the geometry of the binary-lens model.
The straight line with an arrow is the source trajectory, 
red closed concave curves represent the caustics, and 
blue filled circles (marked by $M_{1}$ and $M_{2}$) 
are the binary-lens components. 
All length scales are normalized by the Einstein radius. 
The inset shows the general view and the major panel shows 
the enlarged view corresponding to the light curve of lower panel. 
The open circle on the source trajectory is the source position 
at the time of observation whose size represents the source size. 
(2) The lower panel shows the enlarged view of the anomaly region. 
The inset shows a zoom of the light curve near ${\rm HJD}' \sim 7501.4$. 
The curve superposed on the data is the best-fit binary-lens model.
}
\end{figure}

In Table~\ref{table:one}, we present the best-fit parameters 
of the binary-lens solution along with the $\chi^{2}$ values per 
degree of freedom (dof). The lens system induces two sets of caustics, 
where one is located near the barycenter of the binary lens and the 
other is located away from the barycenter between the planet and the host. 
In Figure~\ref{fig:three}, we show the light curve of the binary-lens solution 
in the region of the anomaly and the corresponding geometry of the lens system. 
We note that the source trajectory in the upper panel is aligned so that 
the data points marked on the trajectory match those in the lower panel.

The fit in Figure~\ref{fig:three} appears quite good. However, 
there exists some residuals near the exit of the planetary caustic 
$7517 < {\rm HJD'} < 7520$. In order to fit the residuals, we 
additionally test the models considering both the microlens parallax 
\citep{gould92a,gould04} and the orbital motion of the lens \citep{dominik98}. 
From this, we find that the introduction of higher-order effects 
still does not provide a fully acceptable fit. Furthermore, 
the estimated {\it projected} energy ratio between 
kinetic and potential energy $(\rm KE/\rm PE)_{\bot} = 5.77$, 
in contrast to the requirement that $(\rm KE/\rm PE)_{\bot} < 1.0$ 
to be a bound system \citep{dong09}, indicates that the model is 
physically unrealistic.  

\begin{figure}[th]
\epsscale{1.2}
\plotone{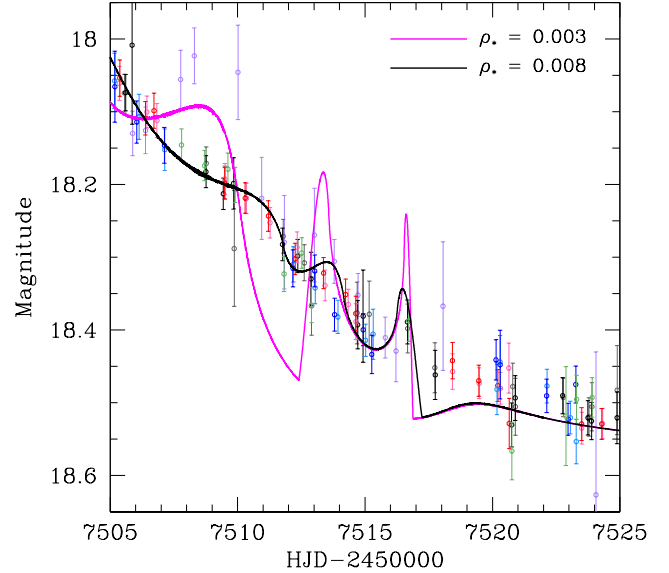}
\caption{\label{fig:four}
Comparison of two binary-lens models in the weak anomaly region.
The black and magenta curves represent, respectively, 
the best-fit model and the same model except with a substantially 
smaller (and more typical) source size $\rho_*$. 
Each data set is binned in time intervals of $24~{\rm hr}$, 
each of which contains a maximum of 10 hrs of contiguous data.    
}
\end{figure}

In fact, there are two important clues that this solution 
is not correct. At first sight these appear to be 
independent, but are actually closely related. The first is 
that the normalized source size is very large $\rho_* = 8\times10^{-3}$ 
considering that the source is a dwarf. This would lead to a very 
small Einstein radius $\theta_{\rm E} \sim 0.05\,$mas and very small 
lens-source relative proper motion $\mu = \theta_{\rm E}/t_{\rm E} 
\sim 0.7\,{\rm mas\,yr^{-1}}$, where $t_{\rm E}$ is the Einstein 
time scale. As discussed in \citet{penny16}, where they presented 
the combination of $\mu$ and $\theta_{\rm E}$ for Galactic 
microlensing events using the model of \citet{henderson14}, 
these values are a priori extremely unlikely, although not impossible: 
they could be generated by a very slow lens at a distance 
$D_{\rm LS} \sim 20\,{\rm pc}(M/M_\odot)^{-1}$ 
in front of the source.

What raises this suspiciously large $\rho_*$ to the level of 
implausibility is that it appears to be ``driven'' by the need 
to match basically smooth data to an intrinsically ``bumpy'' model. 
Because the source crossing time $t_* \equiv \rho_* t_{\rm E}\sim 5\,$hr 
is similar to the duration of an observing night, it is appropriate 
to bin these data by observatory and by day. Figure~\ref{fig:four} compares 
these binned data to the best-fit model and also to the same 
model but with a more typical value of $\rho_* = 3\times10^{-3}$. 
This shows that the data are much smoother than the model for a 
typical $\rho_*$ and are still considerably smoother than the model 
for the best-fit $\rho_*$. We therefore are led to consider 
what other physical effects might generate this ``planetary'' anomaly.

\subsection{Binary-source Interpretation}

We consider the possibility that the long-term deviation may be caused 
by a single-lens event with a binary source. 
In case of a binary-source event, the lensing magnification corresponds 
to the superposition of the two single-source events generated by the 
individual source stars \citep{griest92,han02}, i.e.
\begin{equation}
A = {A_{1}F_{1}+A_{2}F_{2} \over F_{1}+F_{2}} = {A_{1}+q_{F}A_{2} \over 1+q_{F}},  
\label{eq2}
\end{equation}
where $A_{i}$ denotes the lensing magnification of each source star 
with a flux $F_{i}$ and $q_{F} = F_{2}/F_{1}$ is the flux ratio between 
the two source stars. In contrast to a single-source magnification, 
the total magnification $A$ is wavelength dependent 
because the different colors between two sources can induce a color 
change during the course of the lensing phenomenon.
 
\begin{figure}[th]
\epsscale{1.2}
\plotone{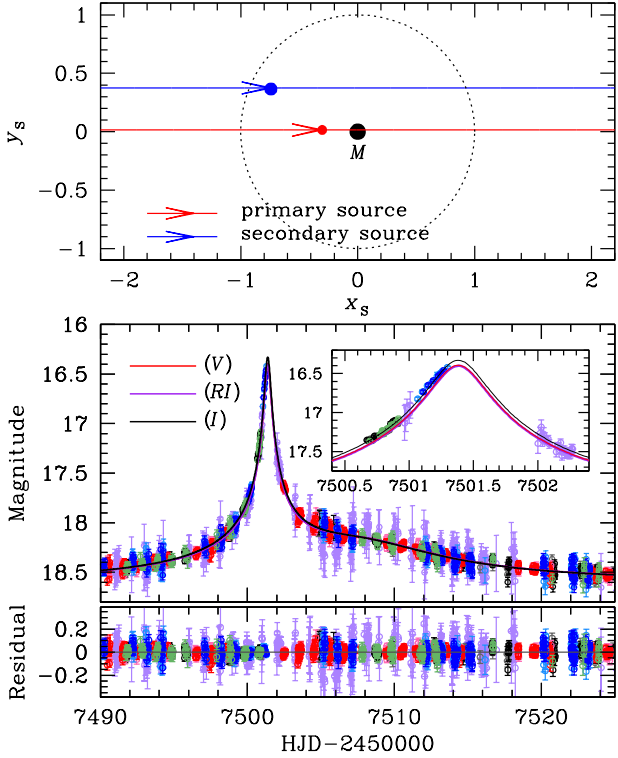}
\caption{\label{fig:five}
Geometry and light curve of the binary-source model. 
(1) The upper panel shows the geometry of the binary-source model. 
Two straight lines with arrows are the source trajectories of 
individual source stars. The lens is located at the origin 
(marked by $M$), and the dotted circle is the angular Einstein ring. 
The red and blue filled circles represent the individual source 
positions at ${\rm HJD}' = 7497$. Lengths are normalized by the Einstein 
radius. (2) The lower panel shows the enlarged view of the anomaly region. 
The two curves with different colors are best-fit binary-source models 
for $RI$ and $I$ passbands. The inset shows a zoom of the anomaly near 
${\rm HJD}' \sim 7501.4$. We note that, although we only use the $V$ band 
data for determining the source type, we also present the $V$ band model 
light-curve to compare the color change between passbands.  
}
\end{figure}

Under the approximation that the transverse speeds of the two sources 
with respect to the lens are same, 
the single-source magnification is related to the lens-source 
separation normalized to the angular Einstein radius, $u_{i}$, by
\begin{equation}
A_{i} = {u_{i}^2+2 \over u_{i} \sqrt{u_{i}^2+4}};\qquad  
u_{i}(t) = \left[\left({t-t_{0,i} \over t_{\rm E}}\right)^{2} + u_{0,i}^{2}\right]^{1/2},
\label{eq3}
\end{equation}
where $t_{0,i}$ is the moment of the closest lens approach to each source star, 
$u_{0,i}$ denotes the separation between the lens and individual sources at $t_{0,i}$, 
and $t_{\rm E}$ is the timescale required for the source to cross the angular 
Einstein radius (Einstein time scale). Therefore, one needs 6 principal parameters 
$(t_{0,1}, t_{0,2}, u_{0,1}, u_{0,2}, t_{\rm E}, q_{F})$ 
for the description of a standard binary-source event. Since the flux ratio varies 
depending on the passband, one must add additional flux ratio parameters 
for each of the observed passbands. We note that, due to the unknown sign of $u_{0,i}$, 
the binary-source interpretation suffers a similar degeneracy to the well-known 
satellite parallax degeneracy \citep{refsdal66,gould94}, resulting in four degenerate 
solutions that are generally denoted by $(+, +)$, $(-, -)$, $(+, -)$, and $(-, +)$, 
where the two signs in the parenthesis indicate the sign of $u_{0,1}$ and $u_{0,2}$, 
respectively. 
\footnote{Because of the symmetry of the lensing magnification, one can 
easily obtain the four solutions of the standard binary-source light curve 
by changing the sign of $u_{0,i}$ once the best-fit solution is found.}

\begin{figure}[th]
\epsscale{1.2}
\plotone{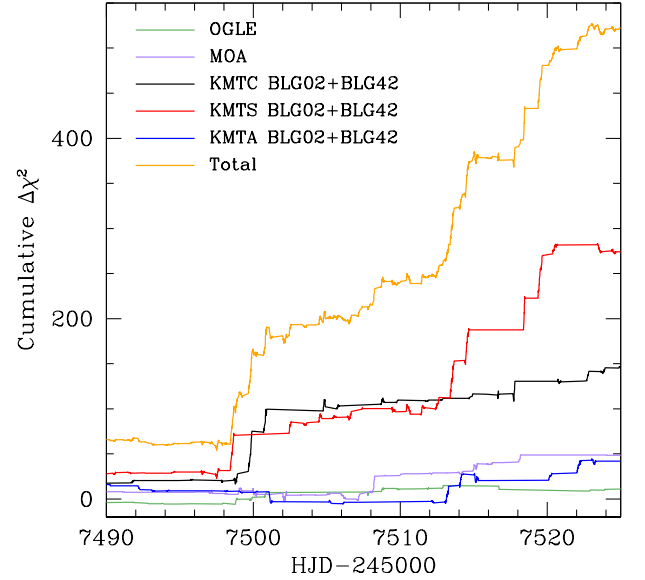}
\caption{\label{fig:six}
Cumulative distribution of $\chi^2$ difference 
between the binary-source and binary-lens models.
}
\end{figure}
The four-fold degeneracy introduces a two-fold degeneracy in the measurement 
of the source plane Einstein radius $\hat{r}_{\rm E}$, defined by 
\begin{equation}
\hat{r}_{\rm E} = D_{\rm S}\theta_{\rm E} = r_{\rm E}{D_{\rm S} \over D_{\rm L}}.
\label{eq4}
\end{equation}
Here $r_{\rm E} = {D_{\rm L}}{\theta_{\rm E}}$ is the physical Einstein radius, 
$D_{\rm S}$ and $D_{\rm L}$ are the distance to the source and the lens, respectively. 
The physical projected separation between the sources, $d_{\rm s}$, is related to 
$\hat{r}_{\rm E}$ by  
\begin{equation} 
d_{\rm s} = {\hat{r}_{\rm E \pm}}\left[\left({\Delta}{t_{0}} \over 
{t_{\rm E}}\right)^{2} + {\Delta}{u_{0}}^{2}\right]^{1/2},
\label{eq5} 
\end{equation} 
where ${\Delta}{t_{0}} = |t_{0,1} - t_{0,2}|$, 
and ${\Delta}{u_{0}} = |u_{0,1} \pm u_{0,2}|$. Therefore, if the separation $d_{\rm s}$ 
is measured from follow-up spectroscopic observation of the sources, 
the Einstein radius is determined with two possible values of 
$\hat{r}_{\rm E \pm}$ depending on ${\Delta}{u_{0}}$ \citep{han97}. 

The transverse speeds of the two sources can be different when the orbital 
motion of the binary source is substantial (``source orbit''). 
In this case, the positions of the individual sources are described 
by the summation of the rectilinear motion of the barycenter of the binary 
source (CM) and the orbital motion of the sources with respect to the CM. 
Similar to the effect of the binary-lens orbital motion \citep{dominik98,jung13}, 
the ``source orbit'' effect also causes the angle between the trajectory of 
the CM and the binary-source axis ${\alpha_{\rm s}}$ and the projected separation 
between the source components $s_{\rm s}$ (normalized to $\theta_{\rm E}$) 
to change in time. To first-order approximation, 
the source-orbital motion can be parameterized 
as $ds_{\rm s}/dt$ and $d{\alpha_{\rm s}}/dt$ which 
represent the change rate of the binary-source separation and 
the orientation angle, respectively. Then, the binary-source 
separation and the orientation angle at time $t$ are
\begin{equation}
s_{\rm s}' = s_{\rm s} + {ds_{\rm s} \over dt} (t - t_{\rm ref}),\qquad
\alpha_{\rm s}' = \alpha_{\rm s} + {d{\alpha_{\rm s}} \over dt} (t - t_{\rm ref}), 
\label{eq6}
\end{equation}
where $t_{\rm ref}$ is the reference time for the binary-source orbital motion. 
The positions of the two sources on the lens plane are thus     
\begin{equation}
u_{i}^{2}(t) = \left[\left({t-t_{0} \over t_{\rm E}}\right) \pm 
s_{\rm s,\it i}\,{\rm cos}\,{\alpha_{\rm s}'}\right]^{2} + 
\left(u_{0} \mp s_{\rm s,\it i}\,{\rm sin}\,{\alpha_{\rm s}'}\right)^{2},
\label{eq7}
\end{equation}
where $s_{\rm s, 1} = s_{\rm s}'q_{\rm s}/(1+q_{\rm s})$ and 
$s_{\rm s, 2} = s_{\rm s}'/(1+q_{\rm s})$ are the separation 
between the two sources and the CM, and 
$q_{\rm s} = M_{\rm s, 2}/M_{\rm s, 1}$ is the mass ratio 
between the source components. Here the parameters 
$(t_{0}, u_{0}, t_{\rm E})$ describe the relative motion 
between the lens and the CM. 
As a result, one requires 9 principal parameters 
$(t_{0}, u_{0}, t_{\rm E}, s_{\rm s}, q_{\rm s}, \alpha_{\rm s}, 
ds_{\rm s}/dt, d\alpha_{\rm s}/dt, q_{F})$ 
to describe the orbital motion of the binary source.

We separately test ``standard'' and ``source orbit'' models. 
In the ``source orbit'' model, we test four degenerate solutions 
resulting from the unknown sign of $u_{0,i}$. We investigate 
the solution of the parameters by a downhill approach. 
The initial values of the parameters are set based on the peak 
time, magnification, and duration of the event. We adopt the 
Markov Chain Monte Carlo (MCMC) method for the $\chi^{2}$ 
minimization of the downhill approach.

From the comparison between the ``standard'' and the ``source orbit'' 
models, it is found that the orbital motion of the binary source does 
not improve the fit significantly. The $\chi^{2}$ difference 
between the ``standard'' and each ``source orbit'' model is very low 
$(\Delta\chi^{2} < 1)$. Consequently, the measured values of two orbital 
parameters $(ds_{\rm s}/dt, d\alpha_{\rm s}/dt)$ are consistent with zero in all 
four ``source orbit'' solutions. These imply that the light curve of the event 
does not have sufficient information for constraining the orbital parameters 
(see the Appendix of \citealt{han16} for a detailed review). We therefore 
only consider the ``standard'' model. 

We find that the binary-source interpretation provides a good fit to the observed 
light curve. In Table~\ref{table:one}, we summarize the best-fit parameters of 
the binary-source solution. In Figure~\ref{fig:five}, we show the geometry of the 
relative lens-source motion (upper panel) and the model light curve (lower panel) 
of the binary-source solution. In the upper panel, the two lines with arrows 
represent the trajectories of the two source stars with respect to the lens 
(marked by ``$M$''). We designate the source passing closer to the lens as 
the ``primary'' source and the other source as the ``secondary'' source. 
We find that the flux ratio ($q_{F,RI} = 2.0$ and $q_{F,I} = 1.9$) is 
greater than unity, indicating that the source approaching closer to the lens 
is fainter than the source approaching farther from the lens. While binary-lens 
light curves are almost perfectly achromatic (after subtraction of the blended 
light), binary-source event light curves can be chromatic, implying that the 
magnification varies depending on the observed passband, and thus we present 
two model light curves corresponding to $RI$ band (MOA) and $I$ band 
(the other groups) data sets. According to the binary-source interpretation, 
the strong anomaly near the peak was produced when the fainter source star 
approached close to the lens and the anomaly in the declining wing was produced 
when the brighter source approached.

\begin{figure}[th]
\epsscale{1.2}
\plotone{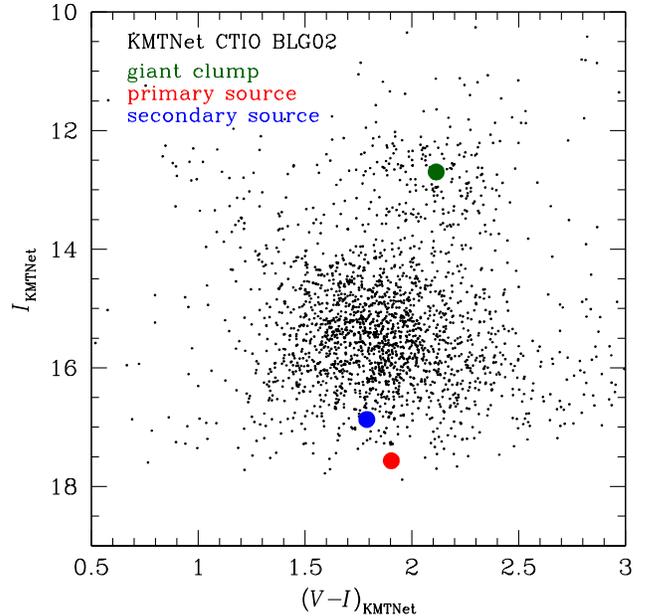}
\caption{\label{fig:seven}
Instrumental color-magnitude diagram of stars in the region 
around the source stars. The positions of individual source 
components and the centroid of the giant clump are marked.  
}
\end{figure}

By comparing the binary-lens and binary-source interpretations, 
we find that the binary-source model better explains the observed 
light curve. The binary-source model improves the fit by 
$\Delta\chi^{2} \sim 544$ compared to the binary-lens model. 
In Figure~\ref{fig:six}, we present the cumulative distribution of 
$\Delta\chi^{2}$ of the binary-source model with respect to the 
binary-lens model as a function of time. This shows that the major 
improvements occur during the ascending part of the strong anomaly 
near $7498 < {\rm HJD'} < 7501$ and the descending part of the light 
curve near $7514 < {\rm HJD'} < 7520$, where the KMTNet survey densely 
and continuously covers the anomaly. The fit improvement from KMTNet 
data is $\Delta\chi^{2} \sim 484$. This demonstrates the importance 
of the high-cadence observation for the unambiguous characterization 
of the lensing event.

\subsection{Source Type}

Knowing that OGLE-2016-BLG-0733 is better explained by the binary-source 
interpretation, we characterize the individual source stars by estimating the 
source type from the de-reddened color $(V-I)_{0}$ and brightness $I_{0}$. 
For this, we use four KMTNet pyDIA reductions ([CTIO+SAAO]$\times$[BLG02+BLG42]) 
to construct the instrumental color-magnitude diagrams (CMD). 
Based on the method of \citet{yoo04}, we first estimate the offsets 
in color $\Delta(V-I)$ and magnitude $\Delta{I}$ 
between individual sources and the centroid of giant clump (GC). 
Figure~\ref{fig:seven} shows the instrumental CMD of one of four KMTNet fields 
(CTIO BLG02) where the locations of individual sources and GC are shown. 
Assuming that the source and GC suffer the same extinction 
combined with the known de-reddened color $(V-I)_{0,\rm GC} = 1.06$ 
and magnitude $I_{0,\rm GC} = 14.43$ of GC \citep{bensby11,nataf13},
we then estimate the de-reddened color and magnitude of the individual 
sources. By applying this procedure to each KMTNet field, we find that 
the mean de-reddened color and magnitude of the individual sources are 
$(V-I, I)_{0,1} = (0.79 \pm 0.10, 19.32 \pm 0.05)$ for the primary 
and $(V-I, I)_{0,2} = (0.68 \pm 0.07, 18.63 \pm 0.03)$ for the secondary. 
These correspond to a late and an early G-type main-sequence star 
for the primary and the secondary, respectively.

Note that we also characterize the source star based on 
the binary-lens interpretation solely for the purpose of 
incorporating the limb-darkening coefficients into the 
binary-lens model. By following the same procedure, 
we determine the mean de-reddened color and magnitude of 
the source as $(V-I, I)_{0} = (0.76 \pm 0.08, 19.60 \pm 0.05)$ 
corresponding to a G-type main-sequence star.

\section{Discussion}

For the great majority of microlensing planets published to date
(see \citealt{mroz16} for a recent listing), the light curves exhibit 
sharp caustic crossings indicative of a binary lens. While in some 
cases there is difficulty in establishing whether the companion 
is planetary or has a more-equal mass ratio, this potential ambiguity 
is decisively resolved in almost all cases.

However, a high-cadence round-the-clock survey 
(such as those reported here) is capable of detecting 
much subtler planetary signals than those published to date. 
Furthermore, such a survey enables a complete statistical analysis 
that addresses all possible planetary signals rather than 
just individual detections. At the same time, \citet{zhu14a,zhu14b} 
have shown that of order half of the planets detectable will lack 
caustic crossing features. Hence, the problem of distinguishing 
such planetary light curves from other physical effects will be 
much more challenging than in the past, but extremely important for 
interpreting statistical results about planet populations.

Here we have presented the first of a new class of potential planet impersonators: 
roughly equal-luminosity binary source events, with the two sources having 
a factor of two flux difference and being separated by substantially 
less than an Einstein radius. This can be regarded as a form of the degeneracy 
proposed by \citet{gaudi98} in the sense that a binary source is masquerading 
as a planet. However, it lies quite far in parameter space from the original 
Gaudi degeneracy from the standpoint both of the physical nature of the binary 
source and the characteristics of the planetary light curve that are being imitated. 
For the Gaudi degeneracy, the second source is much fainter than 
the first and should be separated by of order an Einstein radius or more. 
In this way, the perturbation can appear to be short, and its 
symmetric form will not conflict with asymmetries of planetary 
perturbations generated by planets with projected separations near 
the Einstein ring. By contrast, the binary-source system of 
OGLE-2016-BLG-0733 has roughly equal brightness and so 
induces broad perturbations to the light curve, 
which are then confused with a broadened central peak 
and large planetary deviation, which are characteristic of a planet 
near the Einstein ring.

We found that, once binary-source models were investigated, 
it was quite easy to see that it is preferred over the binary-lens model 
with $\Delta\chi^2 \sim 500$. Moreover, the planetary model exhibited 
``suspicious'' behavior, which is what led us to investigate the binary 
source model. Thus, it may seem that there is no reason to worry that 
this could be an extension of the Gaudi degeneracy (or any kind of degeneracy 
at all). However, unambiguously distinguishing between these two classes of 
models was dependent on having high-cadence data, which may not be available 
in all cases. Furthermore, in the future, a blind search for planetary signals 
will not turn up all potential competing models unless those models are 
specifically considered. The microlensing event OGLE-2016-BLG-0733 
demonstrates another, previously unknown, potential ambiguity that 
could confuse systematic analyses of microlensing planetary signals.

\acknowledgments
OGLE project has received funding from the National Science Centre, Poland, 
grant MAESTRO 2014/14/A/ST9/00121 to AU. The MOA project is supported by 
JSPS KAKENHI Grant Number JSPS24253004, JSPS26247023, JSPS23340064 and 
JSPS15H00781. C. Han acknowledges support from Creative Research 
Initiative Program (2009-0081561) of National Research Foundation of Korea. 
A. Gould is supported from NSF grant AST-1516842 and Korea Astronomy and Space Science 
Institute (KASI) grant 2016-1-832-01. The KMTNet telescopes are operated 
by the Korea Astronomy and Space Science Institute (KASI).

\end{document}